\documentclass[prc,aps,twocolumn,showpacs]{revtex4}
\usepackage{graphicx}
\newcommand{\be}{\begin{eqnarray}}
\newcommand{\ee}{\end{eqnarray}}

\newcommand{\ovl}{\overline}

\begin{document}
    
\title{Resonances, and mechanisms of $\Theta$-production}

\author{Ya.~I.~Azimov$^{1,}$$^2$\footnote[1]{Email: azimov@pa1400.spb.edu},
I.~I.~Strakovsky$^3$\footnote[2]{Email: igor@gwu.edu}
\vspace*{0.1in} 
}

\affiliation{
$^1$Petersburg Nuclear Physics Institute, Gatchina,
    St.~Petersburg 188300, Russia \\
$^2$Thomas Jefferson National Accelerator Facility,\\
    Newport News, VA 23606, USA\\
$^3$Center for Nuclear Studies, Department of Physics,
    The George Washington University, Washington, D.C.
    20052, USA \\
}

\begin{abstract}

After explaining necessity of exotic hadrons, we discuss
mechanisms which could determine production of the exotic
$\Theta$-baryon.  A possible important role of resonances
(producing the $\Theta$ in real or virtual decays) is
emphasized for various processes.  Several experimental
directions for studies of such resonances, and the 
$\Theta$ itself, are suggested.  We briefly discuss also 
recent negative results on the $\Theta$-baryon.

\end{abstract}


\maketitle

The problem of multiquark (exotic and/or ``cryptoexotic") 
states is as old as quarks themselves. The first experimental 
results on searches for exotics~\cite{cool,abr,tys} were 
published soon after invention of quarks~\cite{GM,Z}. 
Initial straightforward motivation ``Why not?" was later 
supported by duality considerations~\cite{ros} (the duality 
was understood at those times as correspondence between 
the sum over resonances and the sum over reggeons). However, 
several years of experimental uncertainty generated the 
question: ``Why are there no strongly bound exotic states 
..., like those of two quarks and two antiquarks or four 
quarks and one antiquark?"~\cite{lip}.     

An attempt to give a reasonable, though model-dependent, 
answer to this question was made in the confined relativistic 
quark model (so called MIT bag)~\cite{jj,jtalk,jprd}. Its 
main conclusion was that the multiquark states should exist, 
and so ``... either these states will be found by 
experimentalists or our confined, quark-gluon theory of 
hadrons is as yet lacking in some fundamental, dynamical 
ingredient which will forbid the existence of these states 
or elevate them to much higher masses"~\cite{jj}. 

What is very essential, neither of approaches based on QCD could 
change this statement, which, therefore, has become even stronger
with time going. However, details of expected properties of exotic 
hadrons are rather different in different approaches. For instance, 
the MIT bag prescribes $J^P=1/2^-$ for the lightest baryon with 
$S=+1$~\cite{jtalk}, while the chiral soliton approach (ChSA) 
predicts $J^P=1/2^+$ (see Refs.~\cite{wk,elkp} for recent re-analyses 
of ChSA predictions and more detailed references). Mass of such 
a baryon should be either about 1700~MeV, in MIT bag~\cite{jtalk}, 
or, in ChSA, most probably below 1600~MeV~\cite{pras}.
Predicted widths of exotic hadrons strongly differ as well. MIT bag 
explains unsuccessful searches for exotic states by their too
broad widths, of several hundreds MeV~\cite{jj,jtalk,jprd},
while, according to ChSA, at least some exotic states may be
quite narrow as compared  to familiar resonances~\cite{DPP}.
Numerous more recent theoretical papers use various theoretical
approaches, and yet could not resolve ambiguities for expected
properties of the exotic hadrons.

Long-time absence of definite experimental results on exotics 
had practically stopped the corresponding activity, and Reviews 
of Particle Properties ceased to touch the exotics problem
after the issue of 1986~\cite{RPP86}. Nevertheless, the paper 
of Diakonov, Petrov, and Polyakov~\cite{DPP}, that predicted the
lightest exotic baryon to have mass about 1530~MeV and width less 
than 15~MeV, strongly stimulated new experimental attempts. They 
provided, at last, positive evidence for the baryon $\Theta^+$
with $S=+1$. Its observation has been stated now in more than 10 
publications~\cite{LEPS,DIANA,clas1,saph,clas2,nu,clas3,herm,SVD,
tof,zeus}, and the measured mass about 1540~MeV looks similar to 
expectations of the ChSA. 

However, spin and parity of $\Theta$ are unknown yet, its  
indirectly estimated width of order 1~MeV~\cite{nus,asw,haid,tril} 
seems to be unexpectedly narrow even for ChSA. Moreover, each of 
the existing positive experiments on $\Theta$ has relatively low
statistics (mainly about 40-50 events above background) which 
looks insufficient today. Therefore, even existence of the 
$\Theta^+$ still needs more indisputable proof.

Meanwhile, there have appeared some experimental publications 
which do not see the $\Theta^+$~\cite{BES,HB,Phen}. Really, 
they do not contradict its existence. Indeed, restrictions of
Ref.~\cite{BES} are rather weak (see Appendix, for their more 
detailed discussion), and some features of data of 
Ref.~\cite{HB} still hint for possibility to extract $\Theta^+$.
Ref.~\cite{Phen} gives the most bright illustration of the present 
uncertain status: the Conference talk with ``a statistically 
significant peak" of $\ovl\Theta^-$ has transformed into the
Proceedings contribution with ``no structure" statement. That
is why we will not discuss here other evidences for the
$\Theta$-non-observation, still being at the level of rumors
and/or slides (a long list of them is given, {\it e.g.}, in
Ref.~\cite{karlip}). Nevertheless, we like to note that searches
for $\Theta^+$ even now use very different processes, with
different initial particles and different energies. Amplitudes 
and cross sections of these processes may (and should) contain  
contributions of various quite different mechanisms, and not
all of them produce the $\Theta$. Therefore, some procedures
to separate the mechanisms may be inevitable, before one can
observe the $\Theta^+$, even if it has been produced.   

We wish to emphasize, however, that if the present evidences
for $\Theta$ appeared incorrect, it would not make the
situation easier, since all the old ``damned" questions on 
exotics would immediately revive. Therefore, we take today more 
conservative position, that $\Theta$ does exist, but its production 
in different conditions is governed by different mechanisms, with 
very different intensity. Though we essentially agree with 
suggestions of Karliner and Lipkin~\cite{karlip} how to clarify 
the problem, we think that, first of all, it is especially 
important to reliably confirm existence of $\Theta$ in the 
processes where it has been stated to be seen. The corresponding 
new data are being collected and treated just now by several
collaborations.
 
In the present note, we discuss qualitative features of possible
mechanisms of the $\Theta$-production and suggest some lines of 
investigations to clarify them.

Even the first information on $\Theta^+$ initiated attempts to 
understand how it is produced, and estimate the production 
cross section. If, for definiteness, we consider the photoproduction
processes
\begin{equation} \label{ntoTh}
\gamma + n \to K^- +\Theta^+ 
\end{equation}
and/or
\begin{equation} \label{ptoTh}
\gamma + p \to \overline K^0 + \Theta^+\,
\end{equation}
(and related electroproduction processes, with virtual
photons), then the most evident contributions come from
exchanges by strange mesons ($K$ and $K^*$, first of all) in the
$t$-channel, and by baryons ($\Theta$ and its possible
excitations) in the $u$-channel. There are also $s$-channel
contributions which correspond, first of all, to formation of
various resonances with non-exotic quantum numbers.  

All those exchange contributions decrease with increasing
energy. To understand this, consider, for example, exchanges by
mesons, $K$ and/or $K^*$. At high energies, they should be
reggeized, and their contributions to the amplitudes are
$\sim{s^{\alpha_i(t)}}$, where $\alpha_i(t)$ is the reggeon
trajectory, with $i=K$ and/or $K^*$. Being integrated over
scattering angles, such contributions reveal energy behavior
$\sim{s^{2\alpha_i(0)-1}}$. Known Regge trajectories may be taken, 
with good accuracy, to be linear, 
$$\alpha(t)\approx\alpha(0)+\alpha't\,,$$
with $\alpha'\approx1$~GeV$^{-2}$. Then, for $K$ and $K^*$
exchanges, having $\alpha_K(m_K^2)=0$ and $\alpha_{K^*}(m_{K^*}^2)
=1$, we obtain $2\alpha_K(0)-1\approx-1.5$ and $2\alpha_{K^*}(0) 
-1\approx -0.6$. Therefore, contributions of the both meson
exchanges, and their interference as well, decrease at high 
energies. Note, that the $K^*$ exchange vanishes somewhat slower 
(and, therefore, becomes more essential) at high energies, than 
the $K$ exchange.  Similar conclusions may be obtained for baryon 
exchanges, and also for exchange contributions in other reactions 
of $\Theta$-production. 
     
Thus, exchanges can not determine the $\Theta$-production at high 
energies, though might be essential at some moderate energies. To
check such possibility, we can compare the $\Theta$
photoproduction processes to strangeness photoproduction with
usual, non-exotic hadrons in the final state. Take, for example, 
reactions 
\begin{equation} \label{NtoY}  
\gamma + N \to K + \Lambda (\Sigma)\,.
\end{equation}
They are kinematically similar to reactions (\ref{ntoTh})
and (\ref{ptoTh}), and have the same $t$-channel exchanges.
These processes have been studied experimentally by different
collaborations~\cite{gstr}. Analyses of the data, up to photon
energies $E_{\gamma}$ of several GeV, suggest that important
contributions come not only from exchanges, but also from
various $s$-channel resonances. Similar conclusions seem to be 
true as well for photoproduction of mesons $\eta$~\cite{geta} 
and $\eta'$ (see Ref.~\cite{geta'} and references therein),
which contain $s\ovl s$ pairs. 

By analogy, we expect that the $\Theta$-photoproduction should 
also be essentially determined by contributions of some 
resonances. What could be those resonances? Up to now, we know
only one such candidate, evidenced for by the CLAS 
Collaboration at JLab~\cite{clas3} and corresponding to a 
rather narrow peak in the mass distribution of the system 
$(K^-\Theta^+)$ near 2400~MeV. We will call it $N^*(2400)$. 

Note, however, that the measured spectrum~\cite{clas3} may 
suggest evidence for some other peaks as well.  Moreover, just 
as in the cases of photoproduction of kaon-hyperon or $\eta$,
and especially for $\eta'$-photoproduction, the resonances
contributing to the $\Theta$-photoproduction do not need to be
real; they can be virtual, subthreshold or overthreshold. So,
even some well-known, rather light nucleon resonances could
participate in reactions (\ref{ntoTh}) and (\ref{ptoTh}), even
though, because of low mass, they can decay to $\ovl K\Theta $
only virtually.  

Resonances may be essential also for the inclusive
$\Theta$-production at high energies. For example, $N^*(2400)$
(or some its analog) might be produced in diffraction dissociation 
of the initial nucleon, and then decay to $\Theta^+$. The 
corresponding cross section could be non-decreasing (or slowly
decreasing) with energy growing. This does not mean that the cross 
section would be large. Just opposite, it will inevitably contain 
a smallness. If the resonance is mainly 3-quark system, its
branching to $\Theta^+$ should be small (we consider the
smallness of the coupling between the $\Theta$ and $KN$ channel
as a general phenomenon). If the resonance is mainly
multi-quark, its branching to $\Theta$ may be large, but its
diffraction production should be suppressed. Thus, the
$\Theta$-production at high energies can be nonvanishing, but
may be essentially determined by other mechanisms, and appear
smaller, as compared to intermediate energies.   

Here we would like to note Ref.~\cite{land} which mainly
reviews results of the SPHINX Collaboration. Its Figs.~5, 11,
and 14a show small, but rather clear bump in the spectrum of
the diffraction excitation $$p\to\Sigma^0K^+\,,$$ having just
$M=2400$~MeV. The same bump seems to be seen at Fig.~12 for 
the excitation $$p\to\Sigma^+K^0\,,$$ and at Fig.~14b for
$$p\to p\,\eta\,.$$  It could be one more independent 
manifestation of $N^*(2400)$. If so, its smallness could be a
confirmation of its (mainly) multi-quark structure.

Since the $N^*(2400)$ is today the only hypothetical resonance
directly related to $\Theta^+$, let us discuss its properties
in some more detail. Isospin of $N^*(2400)$ should be $I=1/2$, 
to allow the decay into $\ovl K\Theta$, with $\Theta$ being 
isosinglet. Further, the state $N^*(2400)$ was 
discovered~\cite{clas3} in the reaction
\begin{equation}
\gamma + p \to \pi^+ + K^- + \Theta^+\,,
~~~~\Theta^+ \to K^+ + n \,,
\end{equation}
being seen as an intermediate stage of the cascade
\begin{equation}
\gamma + p \to \pi^+ + n^*(2400)\,,
~~~~n^*(2400) \to K^- + \Theta^+ \,.
\end{equation}
The kinematical cuts were applied so to enhance contribution 
of the pion exchange. Therefore, $N^*(2400)$ emerges here as 
a resonance in the process
\begin{equation} \label{pipTh}
\pi^- + p \to K^- + \Theta^+\,,
\end{equation}
with the virtual initial pion. This means that $N^*(2400)$  
needs to have nonvanishing coupling to the $\pi N$-channel. It 
should, thus, have the corresponding decay mode, and appear as 
a resonance in the $\pi N$ interaction. Of course, such a
heavy $\pi N$ resonance may have sufficiently small elastic
branching ratio, capable to make it practically unobservable in 
the elastic $\pi N$ scattering. In any case, no partial wave
analysis of $\pi N$ scattering data in this mass range has
seen $N^*(2400)$ with the total width of more or about 100~MeV
and elasticity of more or about 5\%~\cite{hoel}. 

In this connection, it would be very interesting to study the
reaction (\ref{pipTh}) with the real negative pion. We expect
that the process should reveal a rather narrow enhancement at
about $T_{\pi}=2.45$~GeV. Such investigations would be very
interesting for studies of both $\Theta^+$ and $\pi 
N$-resonances. 

Let us discuss possible $SU(3)_F$ properties of $N^*(2400)$. As 
explained, it should be coupled to both $\pi N$ channel (where 
each particle belongs to the corresponding flavor octet), and
$\ovl K\Theta$ (one octet and one antidecuplet hadrons). Since 
(see, {\it e.g.}, Ref.~\cite{SU3})
\begin{equation}
8\times8=1+8_F+8_D+10+\ovl{10}+27, 
~~~8\times\ovl{10}=8+\ovl{10}+27+\ovl{35}\,,
\end{equation}
then, in the case of the exact $SU(3)_F$ symmetry, $N^*(2400)$
should belong to one of the three flavor multiplets: 8,
$\ovl{10}$, or 27 (of course, the antidecuplet here is not that 
which contains $\Theta^+$). 

Studies of $N^*(2400)$, formed in photoproduction (\ref{ntoTh})
and/or (\ref{ptoTh}) as the $s$-channel resonance at 
$E_{\gamma}\approx2.6$~GeV, could help to discriminate these cases. 
To explain this point, we may use the notion of $U$-spin~\cite{Uspin}. 
It is analogous to the $I$-spin, that is, to the familiar isospin. 
But if the $I$-spin mixes $u$- and $d$-quarks, with $s$-quark 
being singlet, then
the $U$-spin mixes $d$- and $s$-quarks, having the same electric 
charge, with $u$-quark being singlet. Therefore, all members of
any $U$-spin multiplet should have the same electric charge.
This implies, that if $SU(3)_F$ is exact and the photon
interaction with quarks is universal, up to electric charges,
the photon is the $U$-spin singlet, and its absorption can not
change $U$-spin of an initial hadron. 

Now, let us compare ``protons" and ``neutrons" in different
unitary multiplets. The $p$-like component of every octet
(together with $\Sigma^+$) belongs to a $U$-spin doublet,
having $U=1/2$. On the other side, the $n$-like component of
the same octet (together with $\Xi^0$ and a combination of
$\Sigma^0$ and $\Lambda^0$ components) is a member of a
$U$-spin triplet, and has $U=1$. For an antidecuplet, the
$n$-like component also has $U=1$ (together with $\Sigma^0$ and
$\Xi^0$), while the $p$-like component has $U=3/2$ (together with 
$\Theta^+, \Sigma^+$, and $\Xi^+$). Situation for a 27-plet is
more complicated: the $p$-like component (with $I=1/2$) is a
superposition of two parts, with $U=1/2$ and 3/2, while the
$n$-like component (also with $I=1/2$) consists of parts with
$U=1$ and 2 (compare to the photon, being the $U$-spin singlet,
but having isoscalar and isovector parts). 

Note, that the initial hadrons in the reaction (\ref{pipTh}) have 
$U(\pi^-)=U(p)=1/2$, and their total $U$-spin can be either 0 
or 1. On the other side, the final hadrons have $U(K^-)=1/2,\, 
U(\Theta^+)=3/2$, and their admissible $U$-spin is 1 or 2. Thus, 
only $U$-vector part of $n^*(2400)$ could contribute to this 
reaction, if $SU(3)_F$ were exact (even if $n^*(2400)$ is the 
member of a 27-plet).

Now, if we compare photoexcitation of $n^*(2400)$ and 
$p^*(2400)$, correspondingly, on the usual $n$ and $p$,
their relation depends on $SU(3)_F$-properties of $N^*(2400)$.
In particular, if $N^*(2400)$ belongs to an antidecuplet, then
photoexcitation of $p^*(2400)$ is forbidden, for exact $SU(3)_F$. 

Of course, $SU(3)_F$ is violated. And nevertheless, one can 
reasonably expect that the photoexcitation of $N^*(2400)$,
being the member of $\ovl{10}$, goes much more intensively on
the neutron than on the proton. As an example, we can remind 
similar consideration~\cite{max} for photoexcitation of the
nonstrange partner of $\Theta^+$ on the neutron and proton with
accounting for $SU(3)_F$-violation.

Interesting information on the nature of $N^*(2400)$ could 
come from its excitation (observed through decay to $\Theta^+$) 
in electroproduction, {\it i.e.} in reactions (\ref{ntoTh})
and (\ref{ptoTh}) with the virtual photon. If the $N^*(2400)$ 
is mainly 5-quark state, then its coupling to the mainly 3-quark 
nucleon should be small at vanishing photon virtuality $Q^2$. 
However, as we know from DIS-studies, the role of multi-quark 
configurations inside the nucleon becomes more important at 
increasing $Q^2$.  This may provide growing of the effective 
$\gamma^*NN^*(2400)$-coupling, when $Q^2$ rises from zero.
Correspondingly, the electroexcitation of $N^*(2400)$ may 
increase with $Q^2$, at least, in some interval from zero.

There is one more way to study the electromagnetic vertex
$\gamma^*NN^*(2400)$. It is to search for the annihilation
\begin{equation} \label{eeN} 
e^+\,e^- \to\ovl N\, N^*(2400)+ {\rm c.c.}\,.
\end{equation}
This could be done inclusively, by missing mass to the nucleon.
Similar search for $N^*$, with subsequent decay $N^*\to N\pi$,
was recently published by BES Collaboration~\cite{bes2}, but
specifically in the peak of $J/\psi$, where only masses below
2160~MeV are kinematically allowed.  The state $N^*(2400)$
could be produced in decays of $\psi(2S)$, but with a different, 
non-electromagnetic vertex. It would provide, therefore, 
different information than the reaction (\ref{eeN}) in 
continuum.
 
Another possibility is to study the exclusive form of the
process (\ref{eeN}), 
\begin{equation} \label{eeK}
e^+\,e^- \to p\,K_S\,\ovl n\,K^- +{\rm c.c.}\,,
\end{equation}
accounting for the consequent decays $$N^*(2400)\to\Theta^+\ovl
K,~~~\Theta^+\to NK\,.$$ The final state (\ref{eeK}) has also
been studied by BES~\cite{BES}, but only in peaks $J/\psi$
and $\psi(2S)$, where the leading contribution is
non-electromagnetic, while the vertex $\gamma^*NN^*(2400)$
appears to be a small correction. It could be essential for
$e^+e^-$-annihilation in continuum, but the present statistics
there is small.       

In summary, we have reminded necessity, at the present level 
of understanding strong interactions, of exotic hadrons, and 
discussed various mechanisms of $\Theta$-production. We have
emphasized, in such processes, a special possible role of
resonances as intermediate objects. Production of $\Theta$ in
very different processes, {\it e.g.}, photo- and 
electroproduction, $e^+e^-$-annihilation, diffraction 
excitation, and others, may be useful to study both the
$\Theta$ itself, and the related resonances.  

\acknowledgments

The authors thank B.~Wojtsekhowski for initiating discussions.
The work was partly supported by the U.~S.~Department of Energy
Grant DE--FG02--99ER41110, by the Jefferson Laboratory, by the 
Southeastern Universities Research Association under DOE
Contract DE--AC05--84ER40150, and by the Russian State Grant
SS--1124.2003.2. Ya.A. acknowledges also the partial support of
Center for Nuclear Studies of the George Washington University.

\appendix
\section{$\Theta^+$ in decays of charmonium}

Collaboration BES investigated decays 
\begin{equation}
J/\psi,\,\psi(2S)\to p\,K_S\,\ovl n\,K^- +{\rm c.c.}
\end{equation}
to search for single and/or double production of $\Theta^+$.
According to their publication~\cite{BES}, $\Theta$ (or
$\ovl\Theta$) was not found at the level of $10^{-5}$. Let us
discuss this in more detail. 

The boundary obtained for the double $\Theta$-production from 
the $J/\psi$ is
\begin{equation}
{\rm Br}(J/\psi\to\Theta\ovl\Theta\to K_SpK^-\ovl n 
+K_S\ovl pK^+n) < 1.1\cdot10^{-5}\,,
\end{equation}
while in the $\psi(2S)$-decays
\begin{equation}
{\rm Br}(\psi(2S)\to\Theta\ovl\Theta\to K_SpK^-\ovl n
+K_S\ovl pK^+n) < 0.84\cdot10^{-5}\,.
\end{equation}
These boundaries can not be directly compared to other known
results. However, using the branching ratios $${\rm
Br}(\Theta\to K^+n)=1/2\,,~~~{\rm Br}(\Theta\to K_Sp)=1/4\,,$$
one can derive
\begin{equation}
{\rm Br}(J/\psi\to\Theta\ovl\Theta) < 0.44\cdot10^{-4}\,,
\end{equation}
\begin{equation}
{\rm Br}(\psi(2S)\to\Theta\ovl\Theta)< 0.34\cdot10^{-4}\,,
\end{equation}
and compare them to other measured branchings. For
instance~\cite{PDG}, $${\rm Br}(J/\psi\to\Lambda
\ovl\Lambda)=(13.0\pm1.2) \cdot10^{-4}\,.$$ At first sight, 
the pair $\Theta\ovl\Theta$ in $J/\psi$-decays is strongly
suppressed in comparison with $\Lambda\ovl\Lambda$, at least,
by the factor \mbox{$<0.034$}. But really, essential part of
this suppression, 0.15, comes from kinematics (\mbox{$S$-wave} 
decay near threshold: c.m. kinetic energy $M_{J/\psi}-
2M_{\Theta}\approx17$~MeV). The dynamical suppression factor is
much weaker, $<0.23\,.$ For decays of $\psi(2S)$, similar
comparison with~\cite{PDG} $${\rm
Br}(\psi(2S)\to\Lambda\ovl\Lambda)=(1.81\pm0.34)\cdot10^{-4}$$ 
gives even weaker suppression, $<0.19\,,$ with the kinematical
factor 0.69 and the dynamical suppression $<0.27$ (compare it
to the dynamical factor $<0.23$ above).

The most stringent restrictions for single $\Theta$-production
are
\begin{equation}
{\rm Br}(J/\psi\to K_Sp\,\ovl\Theta\to K_SpK^-\ovl n)
< 1.1\cdot10^{-5}
\end{equation}
for $J/\psi$ decays, and
\begin{equation}
{\rm Br}(\psi(2S)\to K_Sp\,\ovl\Theta\to K_SpK^-\ovl n)
< 0.60\cdot10^{-5}
\end{equation}
for $\psi(2S)\,.$ Again, one should use branchings to obtain
\begin{equation}
{\rm Br}(J/\psi\to K^0p\,\ovl\Theta) < 0.44\cdot10^{-4}\,,
\end{equation}
\begin{equation}
{\rm Br}(\psi(2S)\to K^0p\,\ovl\Theta)< 0.24 \cdot10^{-4}\,.
\end{equation}
The first of these boundaries may be compared to~~\cite{PDG}
\begin{equation}
{\rm Br}(J/\psi\to K^-p\,\ovl\Lambda)
=(8.9\pm1.6)\cdot10^{-4}\,,
\end{equation}
with the suppression factor $<0.049$. An only appropriate
reference value for decays of $\psi(2S)$ might be~\cite{PDG}  
\begin{equation}
{\rm Br}(\psi(2S)\to\pi^0p\,\ovl p)
=(1.4\pm0.5)\cdot10^{-4}\,,
\end{equation}
which provides the suppression factor $<0.029$. We see that the
total suppression for the single $\Theta$-production in
decays of $J/\psi$ and $\psi(2S)$ is nearly the same as for the
double $\Theta$-production in decays of $J/\psi$ (recall the 
factor of 0.034). It is difficult to separate here kinematical
and dynamical factors, but one can expect somewhat stronger
kinematical suppression in single $\Theta$-decays, because of
3-body phase space. 

Thus, data of BES~\cite{BES} require some suppression in
charmonium decays producing one or two $\Theta$-baryon(s).
However, they still admit rather soft dynamical suppression,
say, 1/5 in the probability. Meanwhile, because of necessity to
produce directly two more quark-antiquark pairs (in exotic
decays as compared with decays to canonical baryon-antibaryon
pairs), some dynamical suppression should naturally arise. It
could be even stronger than the achieved boundaries. Thus, the
recent result of BES~\cite{BES} is only a starting point for
investigating exotics in $e^+e^-$-annihilation.


\end{document}